\documentclass{llncs}%
\usepackage{amsmath}
\usepackage{amsfonts}
\usepackage{amssymb}
\usepackage{graphicx}%
\institute{University of Colorado Denver,
Denver, CO 80217-3364, USA
\\{\tt \{jon.beezley.math,jan.mandel\}@gmail.com}
\and National Center for Atmospheric Research, Boulder, CO 80307-3000, USA
}
\begin{document}

\title{An Ensemble {K}alman-Particle Predictor-Corrector Filter for Non-Gaussian Data Assimilation}
\author{Jan Mandel\inst{1,2}
\and Jonathan D. Beezley\inst{1,2} }
\maketitle

\begin{abstract}
An Ensemble Kalman Filter (EnKF, the predictor) is used make a large change in
the state, followed by a Particle Filer (PF, the corrector), which assigns
importance weights to describe a non-Gaussian distribution. The importance
weights are obtained by nonparametric density estimation. It is demonstrated
on several numerical examples that the new predictor-corrector filter combines
the advantages of the EnKF and the PF and that it is suitable for high
dimensional states which are discretizations of solutions of partial
differential equations.


\end{abstract}

\section{Introduction}

\label{sec:introduction}

Dynamic Data Driven Application Systems (DDDAS) \cite{Darema-2004-DDD} aim to
integrate data acquisition, modeling, and measurement steering into one
dynamic system. Data assimilation is a statistical technique to modify model
state in response to data and an important component of the DDDAS approach.
Models are generally discretizations of partial differential equations and
they may have easily millions of degrees of freedom. The model equations
themselves are posed in functional spaces, which are infinitely dimensional.
Because of nonlinearities, the probability distribution of the state is
usually non-Gaussian.

A number of methods for data assimilation exist \cite{Kalnay-AMD-2003}.
Filters attempt to find the best estimate from the model state and the data up
to the present. We present a combination of the Ensemble Kalman Filter (EnKF)
\cite{Burgers-1998-ASE} and the Sequential Importal Sampling (SIS) particle
filter (PF) \cite{Doucet-2001-SMC}. The EnKF is a Monte-Carlo implementation
of the Kalman Filter (KF). The KF is an exact method for Gaussian
distributions. However, it needs to maintain the state covariance matrix,
which is not possible for large state dimension. The EnKF and its variants
\cite{Anderson-2001-EAK,Mitchell-2000-AEK} replace the covariance by the
sample covariance computed from an ensemble of simulations. Each ensemble
member is advanced in time by the model independently until analysis time,
when the data is injected, resulting in changes in the states of the ensemble
members. Particle filters also evolve a ensemble of simulations, but they
assign to each ensemble member a weight and the analysis step updates the weights.

The KF and the EnKF represent the probability distributions by the mean and
the covariance, and so they assume that the distributions are Gaussian. This
shows in the tendency of EnKF to smear distributions towards unimodal, as
illustrated in Sec. \ref{sec:numerical} below. So, while the EnKF has the
advantage that it can make large charges in the state and the ensemble can
represent an arbitrary distribution, the EnKF is still essentialy limited to
Gaussian distributions. On the other hand, the PF can represent non-Gaussian
distributions faithfully, but it only updates the weights and cannot move
ensemble members in the state space. Thus a method that combines the
advantages of both without the disadvantages of either is of interest. The
design of more efficient non-Gaussian filters for large-scale problems has
been the subject of significant interest, often using Gaussian mixtures and
related approaches \cite{Bengtsson-2003-NFE}.

The predictor-corrector filter presented here uses an EnKF\ as a predictor to
move the state distribution towards the correct region and then a PF as
corrector to adjust for a non-Gaussian character of the distribution.
Nonparametric density estimation is used to compute the weights in the PF. The
combined predictor-corrector method appears to work well on problems where
either EnKF\ of PF fails, and it does not degrade the performace of the EnKF
for Gaussian distributions. Predictor-corrector filters were first formulated
in \cite{Mandel-2006-PEF,Mandel-2007-PME}. Related results and some
probabilistic background can be found in \cite{Mandel-2008-PEF}.

\section{Formulation of the Method}

A\ common procedure to construct an initial ensemble is as a sum with random
coefficients \cite{Evensen-1994-SDA},%

\begin{equation}
u=\sum_{n=1}^{m}\lambda_{n}d_{n}\varphi_{n},\quad d_{n}\sim N\left(
0,1\right)  ,\quad\left\{  d_{n}\right\}  \text{ independent},
\label{eq:random-smooth}%
\end{equation}
where $\left\{  \varphi_{n}\right\}  $ is an orthonormal basis in the space
state $V=\mathbb{R}^{m}$ equipped with the Euclidean norm $\left\Vert
\cdot\right\Vert $. The elements of $V$ are column vectors of values of
functions on a mesh in the spatial domain. The basis functions $\varphi_{n}$
are smooth for small $n$ and more oscillatory for large $n$. If the the
coefficients $\lambda_{n}\rightarrow0$ sufficiently fast, the series
(\ref{eq:random-smooth}) converges and $u$ is a random smooth function in the
limit as $m\rightarrow\infty$. The sum (\ref{eq:random-smooth}) defines a
Gaussian random variable with the eigenvalues of its covariance matrix equal
to $\lambda_{k}^{2}$. Possible choices of $\left\{  \varphi_{k}\right\}  $
include a Fourier basis, such as the sine or cosine functions, or bred vectors
\cite{Kalnay-AMD-2003}. On the state space $V$, we define another norm by
\begin{equation}
\left\Vert u\right\Vert _{U}^{2}=\sum_{n=1}^{m}\frac{1}{\kappa_{n}^{2}}%
c_{n}^{2},\quad u=\sum_{n=1}^{m}c_{n}\varphi_{n}. \label{eq:U-norm}%
\end{equation}
Note that if $\kappa_{n}=1$, $\left\Vert \cdot\right\Vert _{U}$ is just the
original norm $\left\Vert \cdot\right\Vert $ on $V$. We generally use
$\kappa_{n}$ adapted to the smoothness of the functions in the initial
ensemble, $\lambda_{n}/\kappa_{n}\rightarrow0$ as $n\rightarrow\infty$.

A weighted ensemble of $N$ simulations $\left(  u_{k},w_{k}\right)  _{k=1}%
^{N}$ is initialized according to (\ref{eq:random-smooth}), with equal weights
$w_{k}=1/N$. The ensemble members are advanced by the model and at given
points in time, new data is injected by an \emph{analysis step}. The data
consists of vector $d$ of measurements, observation function $h\left(
u\right)  =Hu$, also called forward operator, here assumed to be linear, which
links the model state space with the data space, and data error distribution,
here assumed to be Gaussian with zero mean and known covariance $R$. The value
of the observation function $Hu$ is what the data vector would be in the
absence of model and data errors. The value of the probability density of the
data error distribution at the data vector $d$ for a given value of the
observation function $Hu$ is called \emph{data likelihood} and denoted by
$p(d|u)$. The probability distribution of the model state before the data is
injected is called the \emph{prior} or the \emph{forecast}, and the
distribution after the data is injected is called the \emph{posterior} or the
\emph{analysis}. Assuming the forecast probability distribution has the
density $p^{f}$, the density $p^{a}$ of the analysis is found from the Bayes
theorem,
\begin{equation}
p^{a}\left(  u\right)  \propto p\left(  d|u\right)  p^{f}\left(  u\right)  ,
\label{eq:Bayes-dens}%
\end{equation}
where $\propto$ means proportional.

Instead of working with densities, the probability distributions are
approximated by weighted ensembles. We will call the following analysis step
algorithm \emph{EnKF-SIS}.

\textbf{Predictor.} Given a forecast ensemble
\[
\left(  u_{k}^{f},w_{k}^{f}\right)  _{k=1}^{N},\quad w_{k}^{f}\geq0,\quad
\sum_{k=1}^{N}w_{k}^{f}=1,
\]
the members $u_{k}^{a}$ of the analysis ensemble are found from the EnKF,
\[
u_{k}^{a}=u_{k}^{f}+K(d_{k}-Hu_{k}^{f}),\quad d_{k}\sim N\left(  d,R\right)
,\quad K=QH^{\mathrm{T}}\left(  HQH^{\mathrm{T}}+R\right)  ^{-1}%
\]
where $d_{k}$ are randomly sampled from the data distribution, and $Q$ is the
forecast ensemble covariance,%
\begin{equation}
Q=\sum_{k=1}^{N}w_{k}\left(  u_{k}-\overline{u}^{f}\right)  \left(
u_{k}-\overline{u}^{f}\right)  ^{\mathrm{T}},\quad\overline{u}^{f}=\sum
_{k=1}^{N}w_{k}^{f}u_{k}^{f}.\label{eq:wc}%
\end{equation}
This is the EnKF from \cite{Burgers-1998-ASE}, extended to weighted ensembles
by the use of the weighted sample covariance (\ref{eq:wc}).

\textbf{Corrector.} The analysis members $u_{k}^{a}$ are thought of as a
sample from some \emph{proposal distribution}, with density $p^{p}$. Ideally,
the analysis weights $w_{k}^{a}$ should be computed from the SIS update as
\cite{Doucet-2001-SMC}
\[
w_{k}^{a}\propto p\left(  d | u_{k}^{a}\right)  \frac{ p^{f}\left(  u_{k}^{a}
\right)  } { p^{p} \left(  u_{k}^{a} \right)  }.
\]
However, the ratio of the densities is not known, so it is replaced by a
noparametric estimate inspired by \cite{Loftsgaarden-1965-NEM}, giving
\[
w_{k}^{a}\propto p\left(  d|u_{k}^{a}\right)  \frac{\sum_{\ell:\left\Vert
u_{\ell}^{f}-u_{k}^{a}\right\Vert _{U}\leq h_{k}}w_{k}^{f}}{ \sum
_{\ell:\left\Vert u_{\ell}^{a}-u_{k}^{a}\right\Vert _{U}\leq h_{k}}\frac{1}%
{N}},\quad\sum_{k=1}^{N}w_{k}^{a}=1.
\]
The bandwidth $h_{k}$ is the distance from $u_{k}^{a}$ to the $\lfloor
N^{1/2}\rfloor$-th nearest member $u_{\ell}^{a}$, measured in the $\left\Vert
\cdot\right\Vert _{U}$ norm.

\section{Numerical Results}

\label{sec:numerical}

%\subsection{An illustration in 1D}

\begin{figure}[t]
\begin{center}%
\begin{tabular}
[c]{cc}%
\includegraphics[width=2.1in]{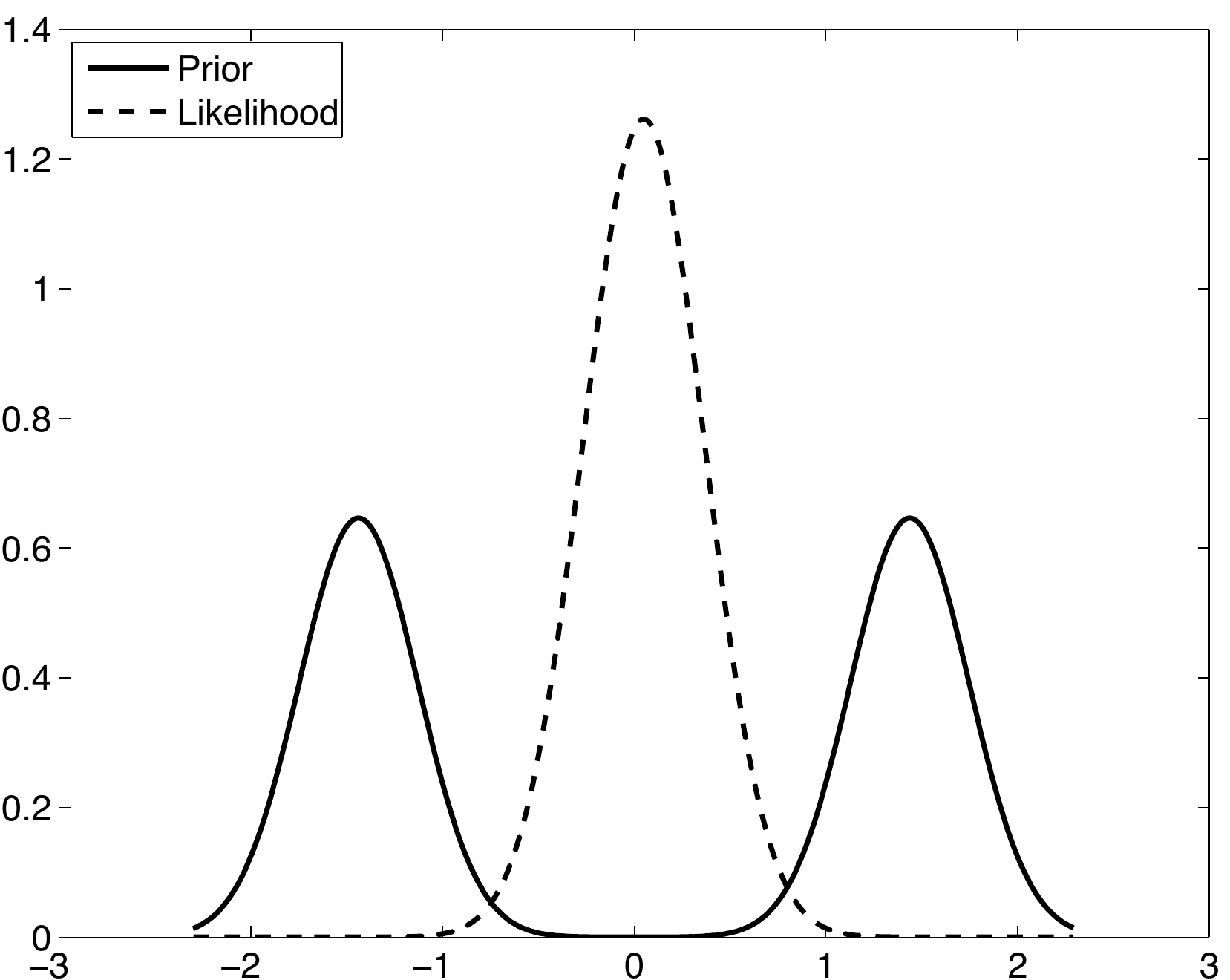} &
\includegraphics[width=2.1in]{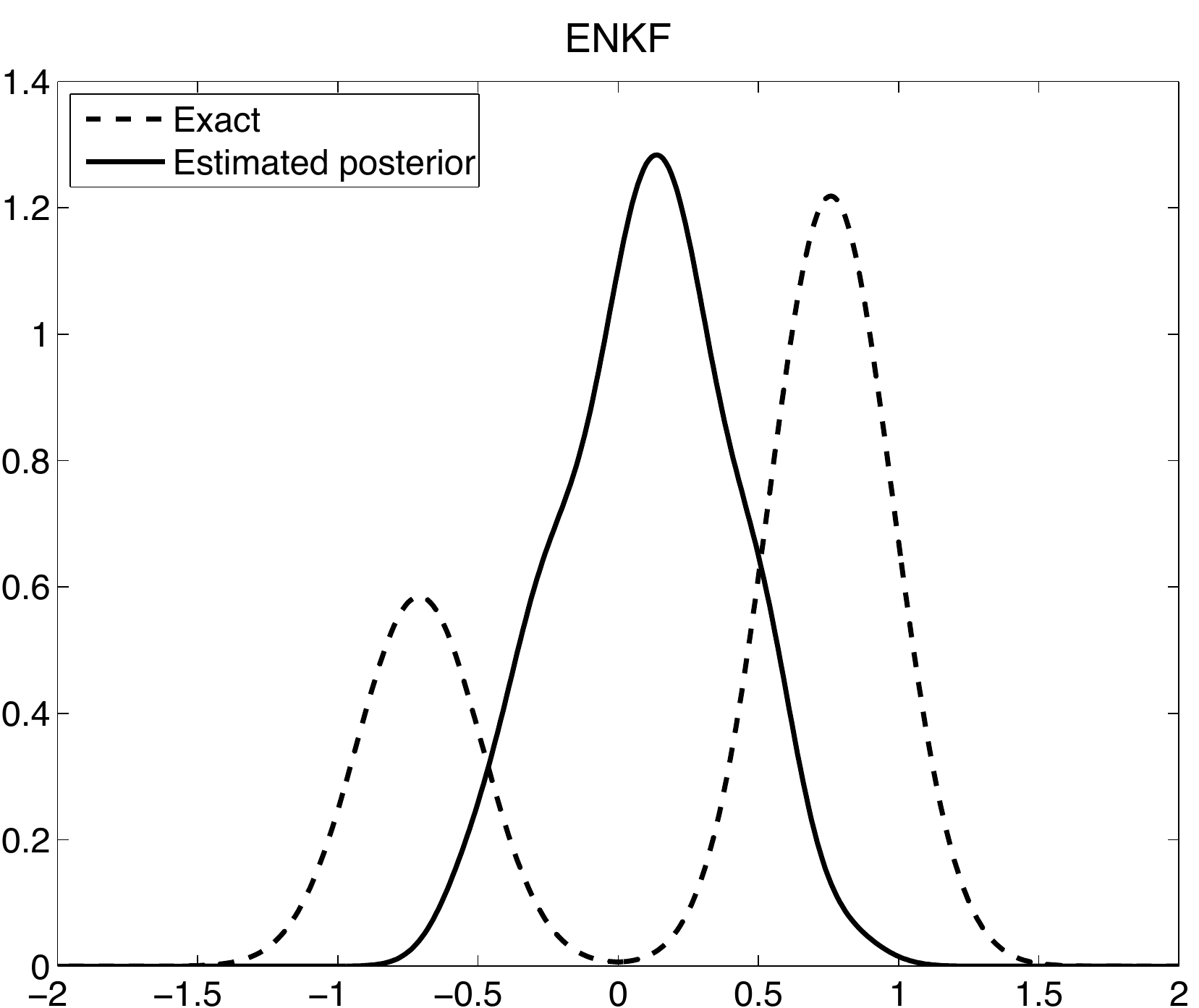}\\
{(a) Prior and data likelihood densities} & {(b) Posterior from EnKF}\\
\includegraphics[width=2.1in]{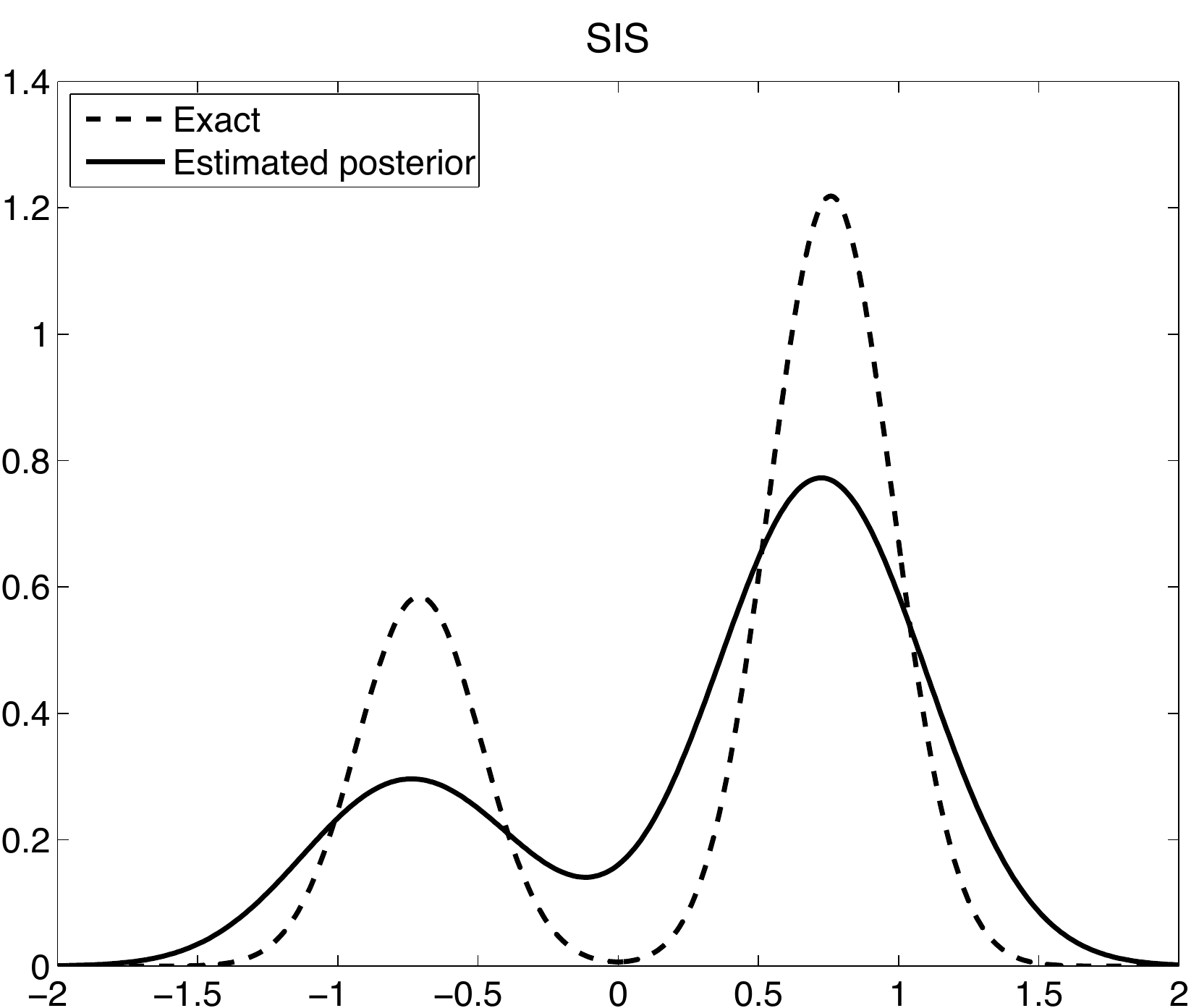} &
\includegraphics[width=2.1in]{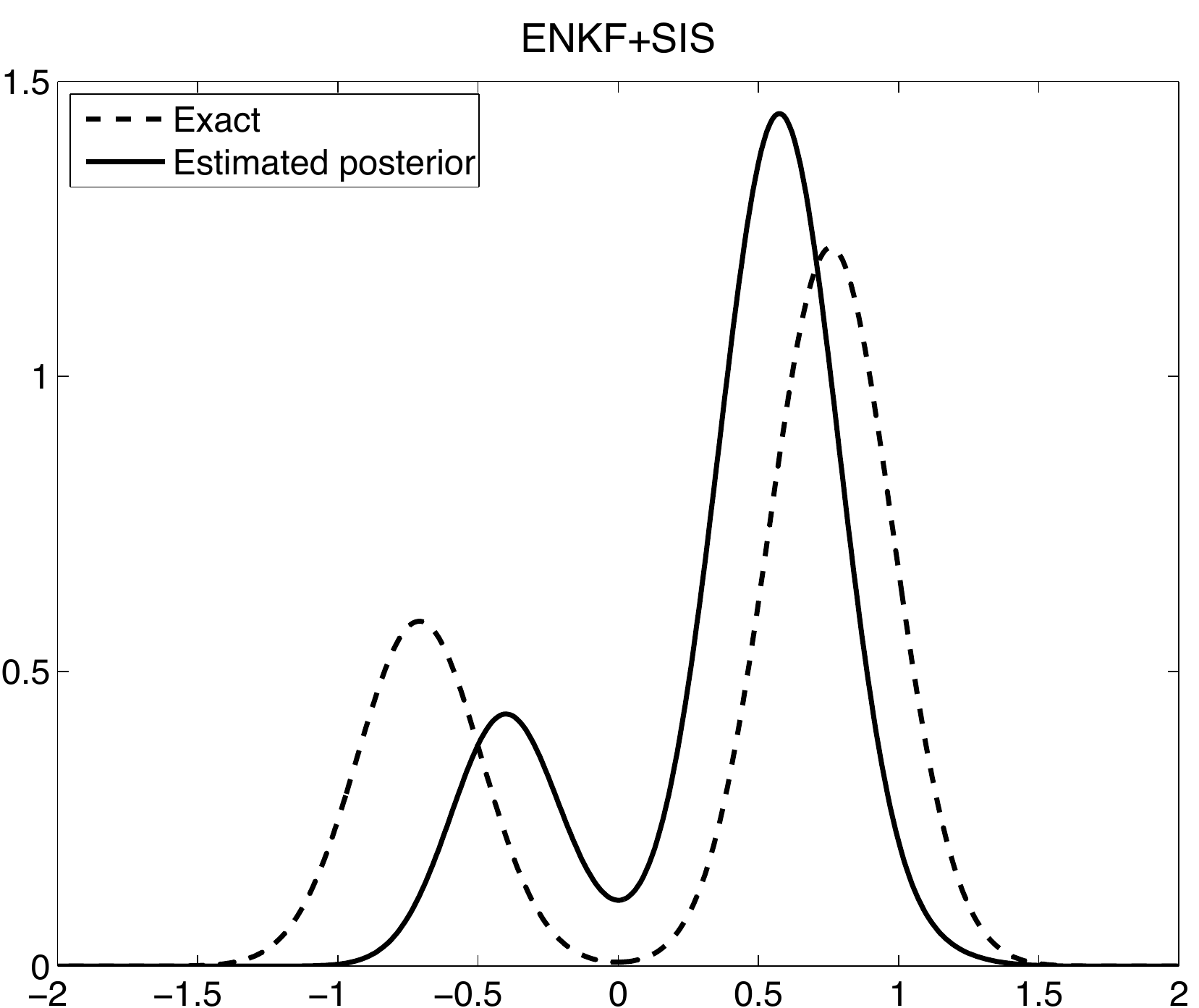}\\
{(c) Posterior from SIS} & {(d) Posterior from EnKF-SIS}%
\end{tabular}
\end{center}
\caption{Data assimilation with bimodal prior. EnKF fails to capture the
non-Gaussian features of the posterior, but both SIS and EnKF-SIS represent
the nature of the posterior reasonably well.}%
\label{fig:bimodal}%
\end{figure}

\begin{figure}[t]
\begin{center}
\hspace*{-0.2in} \includegraphics[width=3in]{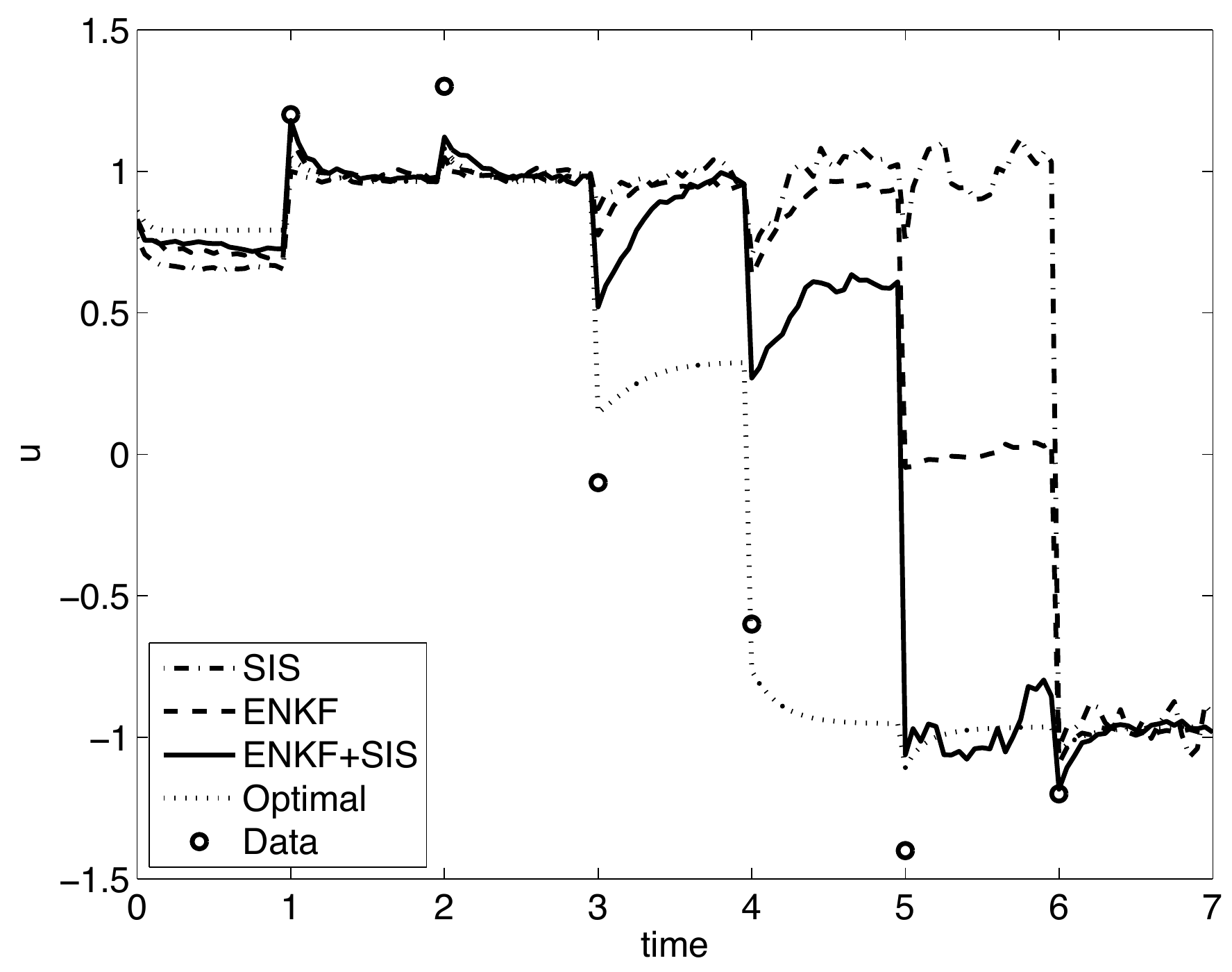}
\hspace*{-0.2in}
\end{center}
\caption{Ensemble filters mean and optimal filter mean for stochastic ODE
(\ref{eq:rsode}). EnKF-SIS was able to approximate the optimal solution better
than either SIS or EnKF alone. }%
\label{fig:rsode_results}%
\end{figure}

Fig.~\ref{fig:bimodal} demonstrates a failure of EnKF for non-Gaussian
distributions, while SIS and EnKF-SIS do fine. We construct a bimodal prior in
1D by first sampling from $N\left(  0,5\right)  $ and assigning the weights
by
\[
w_{f}(x_{i})=e^{-5(1.5-x_{i})^{2}}+e^{-5(-1.5-x_{i})^{2}}.
\]
The data likelihood is Gaussian. The ensemble size was $N=100$.

The next 1D problem demonstrates that EnKF-SIS is doing better that either
EnKF or SIS alone in filtering for the stochastic differential
equation~\cite{Kim-2003-EFN}
\begin{equation}
\frac{du}{dt}=4u-4u^{3}+\kappa\eta, \label{eq:rsode}%
\end{equation}
where $\eta(t)$ is white noise. The parameter $\kappa$ controls the magnitude
of the stochastic term.

The deterministic part of this differential equation is of the form
\[
\frac{du}{dt}=-f^{\prime}\left(  u\right)  ,
\]
where the potential $f(u)=-2u^{2}+u^{4}$. The equilibria are given by
$f^{\prime}\left(  u\right)  =0$; there are two stable equilibria at $u=\pm1$
and an unstable equilibrium at $u=0$. The stochastic term of the differential
equation makes it possible for the state to flip from one stable equilibrium
point to another; however, a sufficiently small $\kappa$ insures that such an
event is rare.

A suitable test for an ensemble filter will be to determine if it can properly
track the model as it transitions from one stable fixed point to the other.
From Fig.~\ref{fig:bimodal}, it is clear that EnKF will not be capable of
describing the bimodal nature of the state distribution so it will not perform
well when tracking the transition. Also, when the ensemble is centered around
one stable point, it is unlikely that some ensemble members would be close to
the other stable point. It is known that SIS can be very slow in tracking the
transition and EnKF can do better \cite{Kim-2003-EFN}.
Fig.~\ref{fig:rsode_results} demonstrates that EnKF can outperform both.

The solution $u$ of (\ref{eq:rsode}) is a random variable dependent on time,
with density $p(t,u)$. The evolution of the density in time is given by the
Fokker-Planck equation, which was solved numerically on a uniform mesh from
$u=-3$ to $u=3$ with the step $\Delta u=0.01$. At the analysis time, the
optimal posterior density was computed by multiplying the probability density
of $u$ by the data likelihood following (\ref{eq:Bayes-dens}) and then scaling
so that again $\int pdu=1$, using numerical quadrature by the trapezoidal
rule. The data points were taken from one solution of this model, called a
reference solution, which exhibits a switch at time $t\approx1.3$. The data
error distribution was normal with the variance taken to be $0.1$ at each
point. To advance the ensemble members and the reference solution, we have
solved (\ref{eq:rsode}) by the explicit Euler method with a random
perturbation from $N(0,\left(  \Delta t\right)  ^{1/2})$ added to the right
hand side in every step \cite{Higham-2001-AIN}. The simulation was run for
each method with ensemble size $100$, and assimilation performed for each data point.

%\subsection{Filtering in high dimension}

\label{sec:hd}

\begin{figure}[t]
\begin{center}%
\begin{tabular}
[c]{cc}%
\includegraphics[width=2.1in]{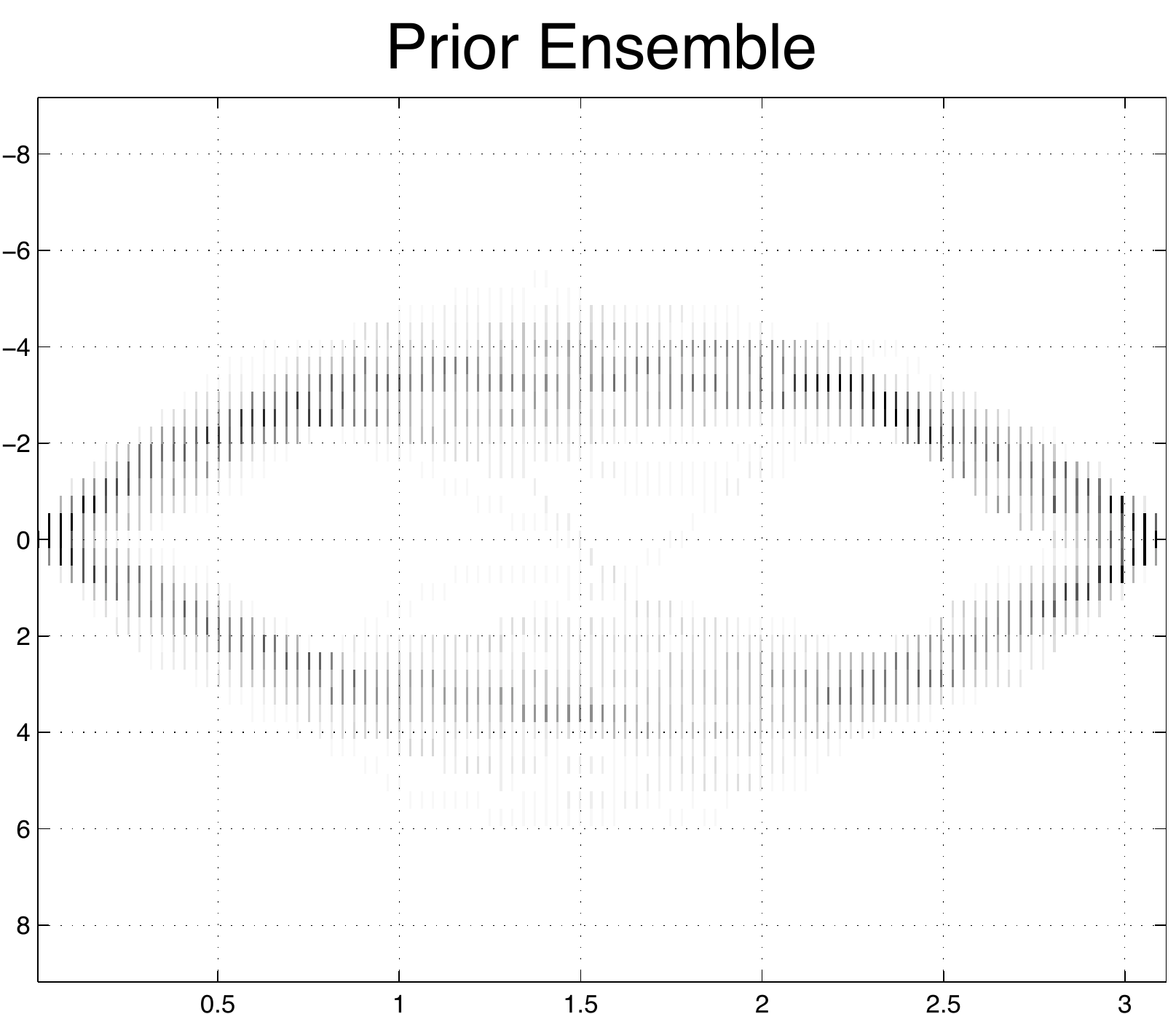} &
\includegraphics[width=2.1in]{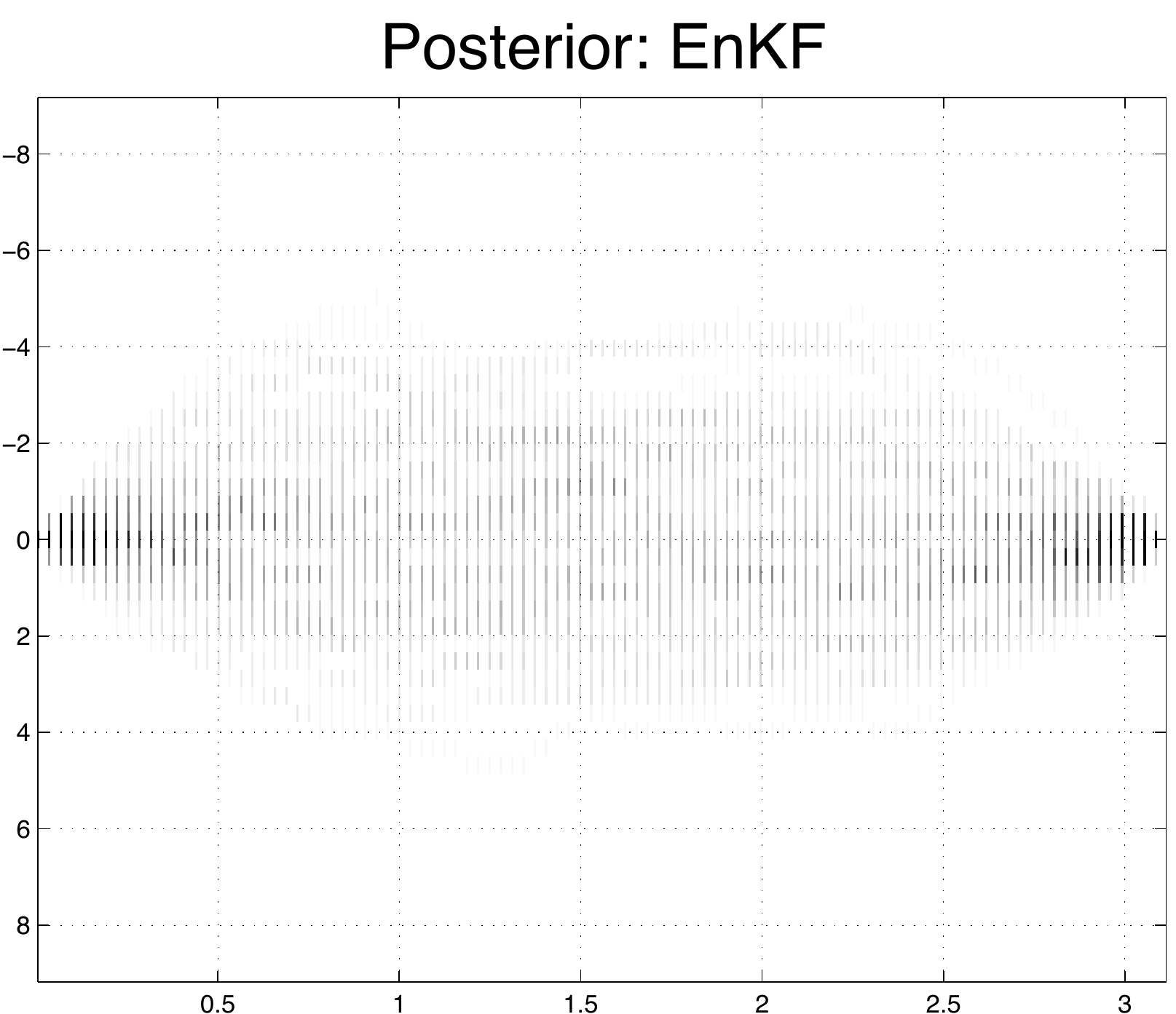}\\
{(a)} & {(b)}\\
\includegraphics[width=2.1in]{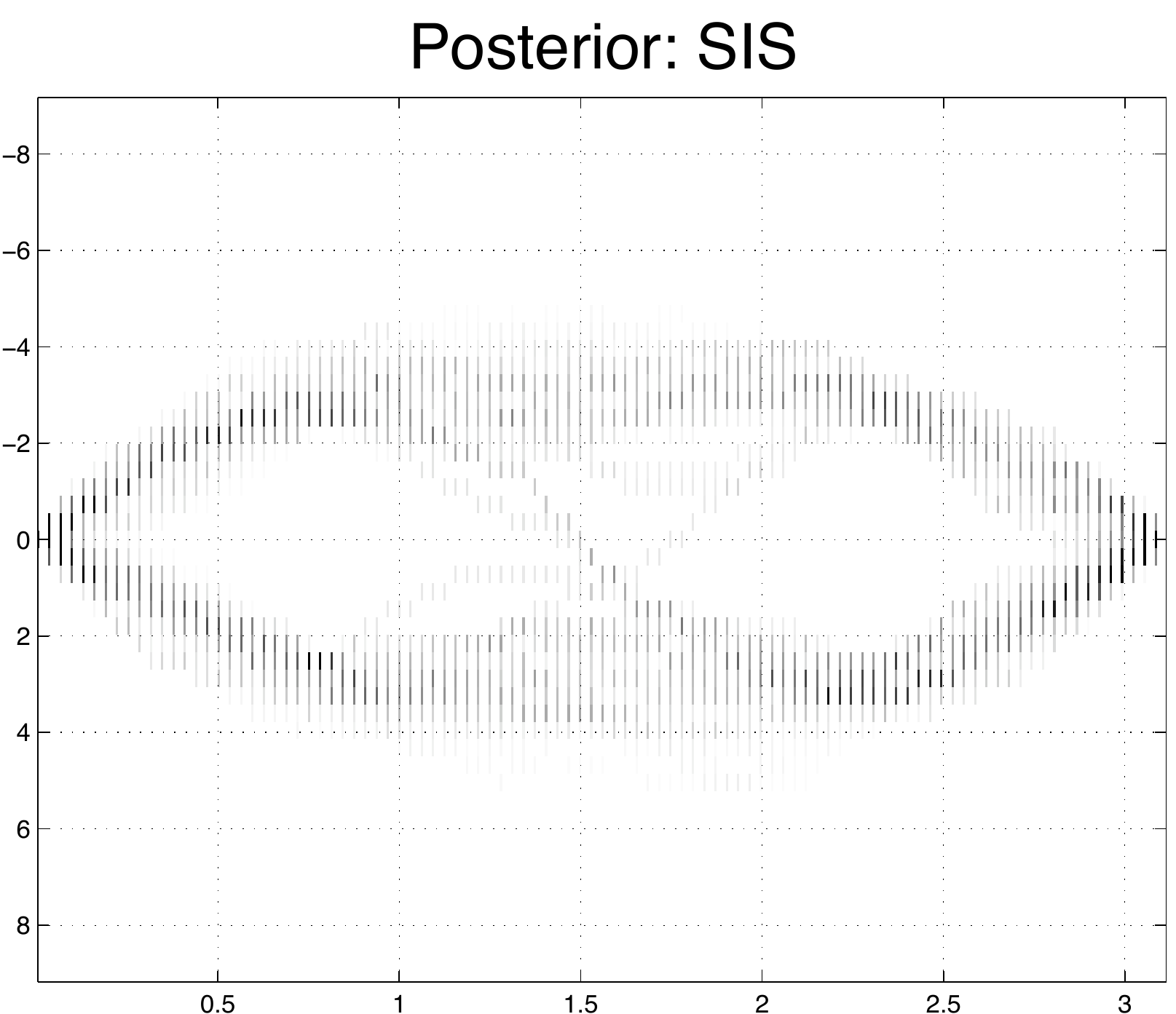} &
\includegraphics[width=2.1in]{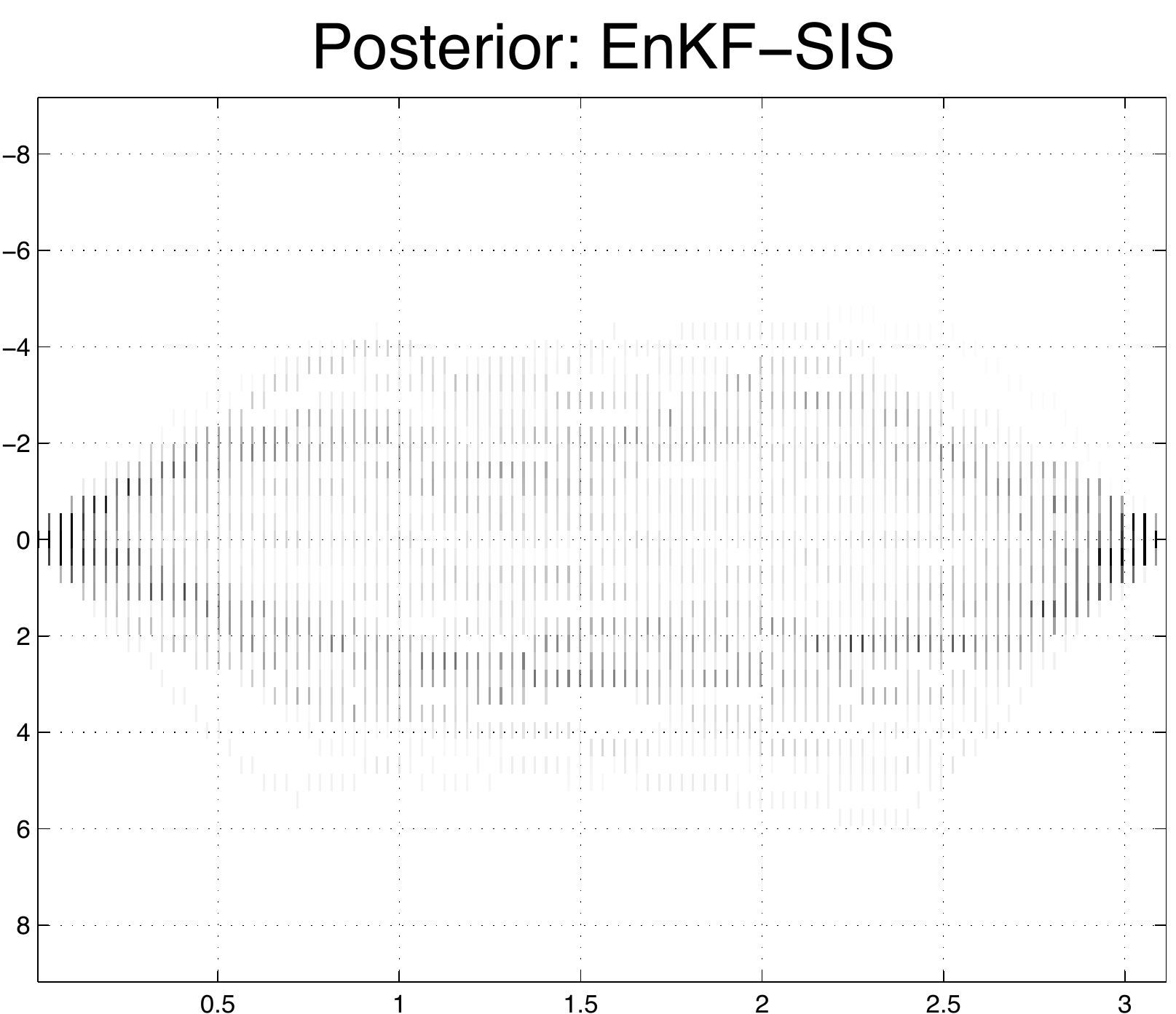}\\
{(c)} & {(d)}%
\end{tabular}
\end{center}
\caption{EnKF smears non-Gaussian distribution. The horizontal axis is the
spatial coordinate $x\in\lbrack0,\pi]$. The vertical axis is the value of
function $u$. The level of shading on each vertical line is the marginal
density of $u$ at a fixed $x$, computed from a histogram with 50 bins. While
EnKF completely ignores the non-Gaussian character of the posterior and
centers the distribution around $u=0$, EnKF-SIS shows darker bands at the
edges. }%
\label{fig:b-enkf}%
\end{figure}

\begin{figure}[t]
\begin{center}%
\begin{tabular}
[c]{cc}%
\includegraphics[width=2.1in]{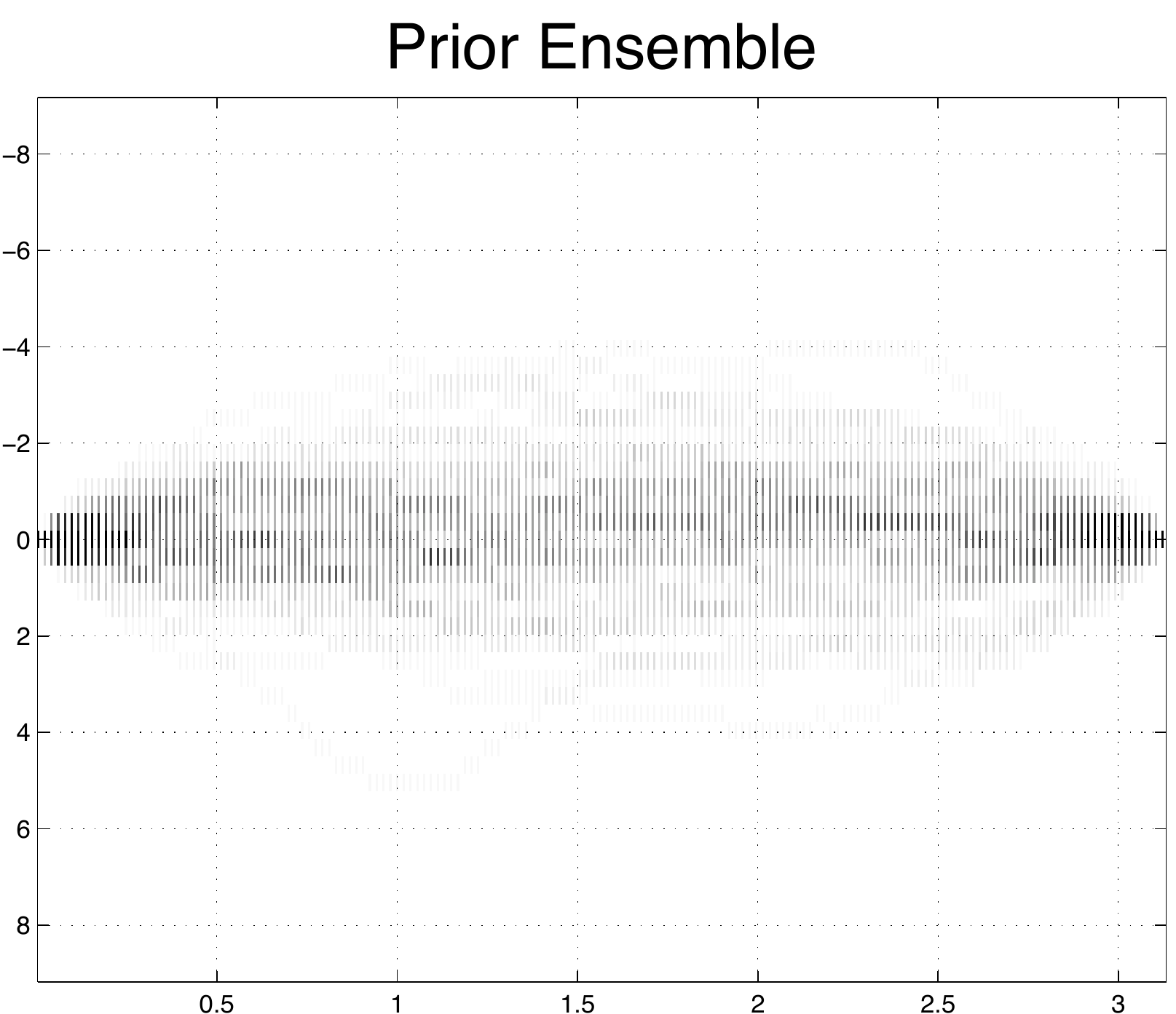} &
\includegraphics[width=2.1in]{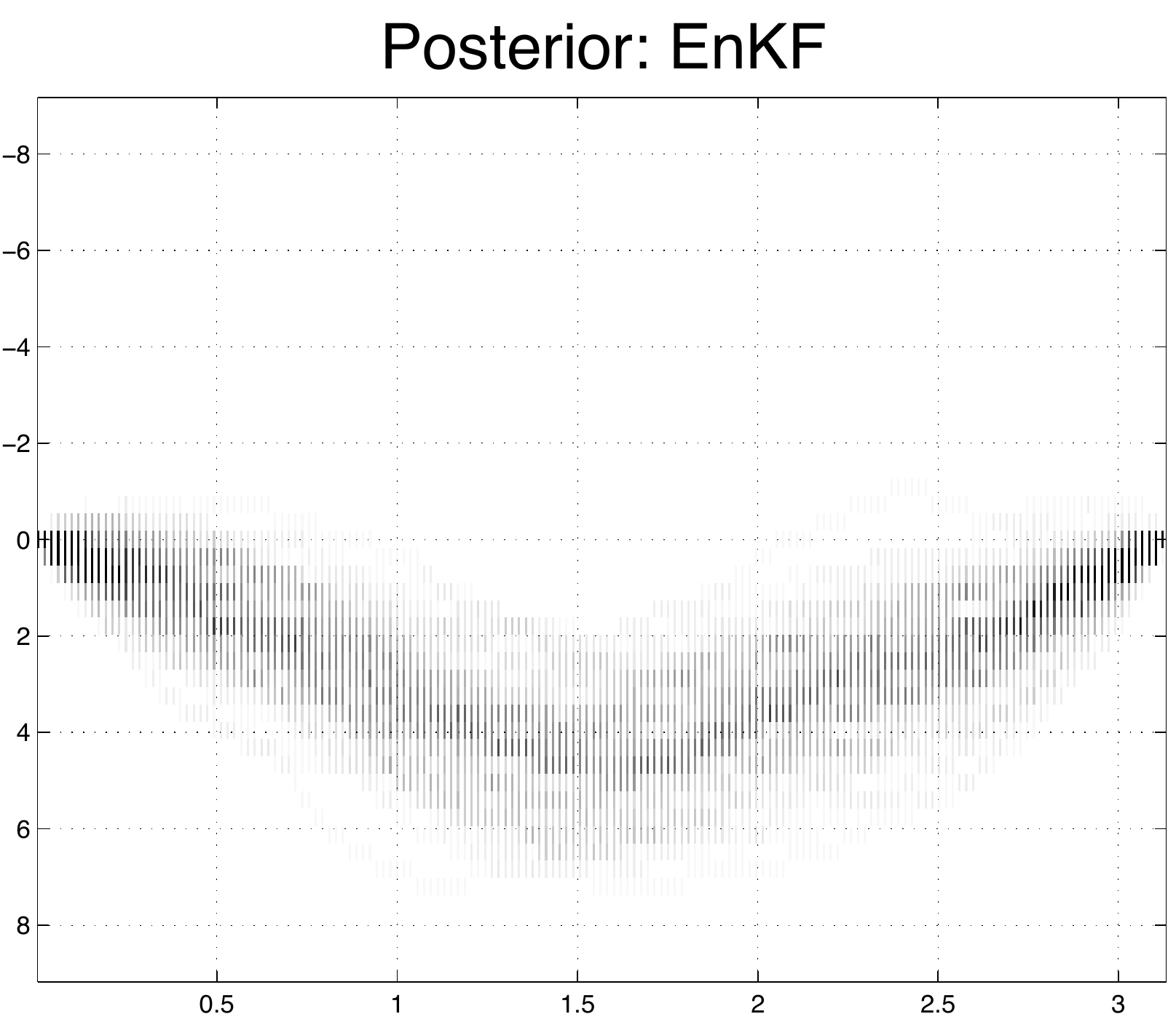}\\
{(a)} & {(b)}\\
\includegraphics[width=2.1in]{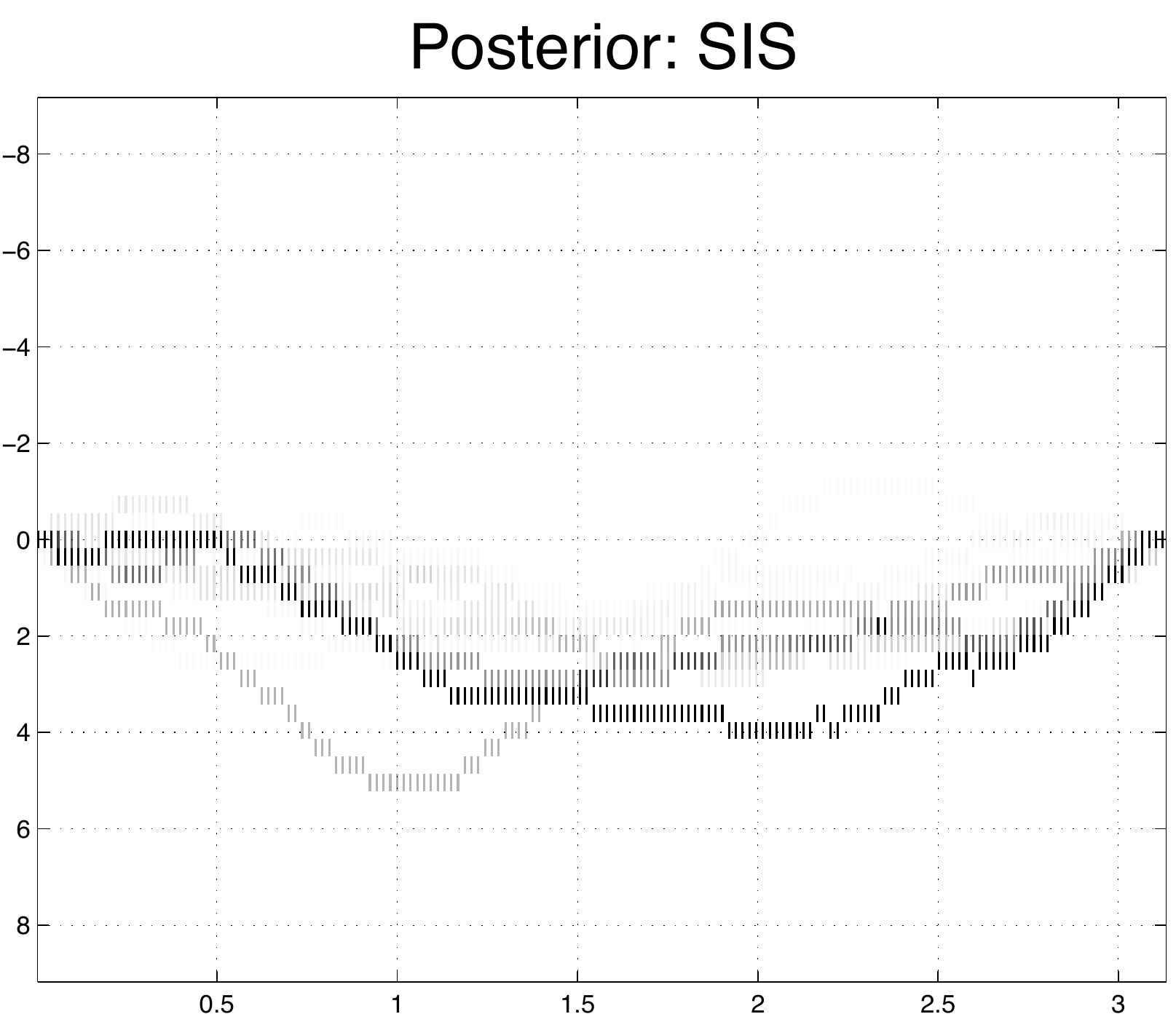} &
\includegraphics[width=2.1in]{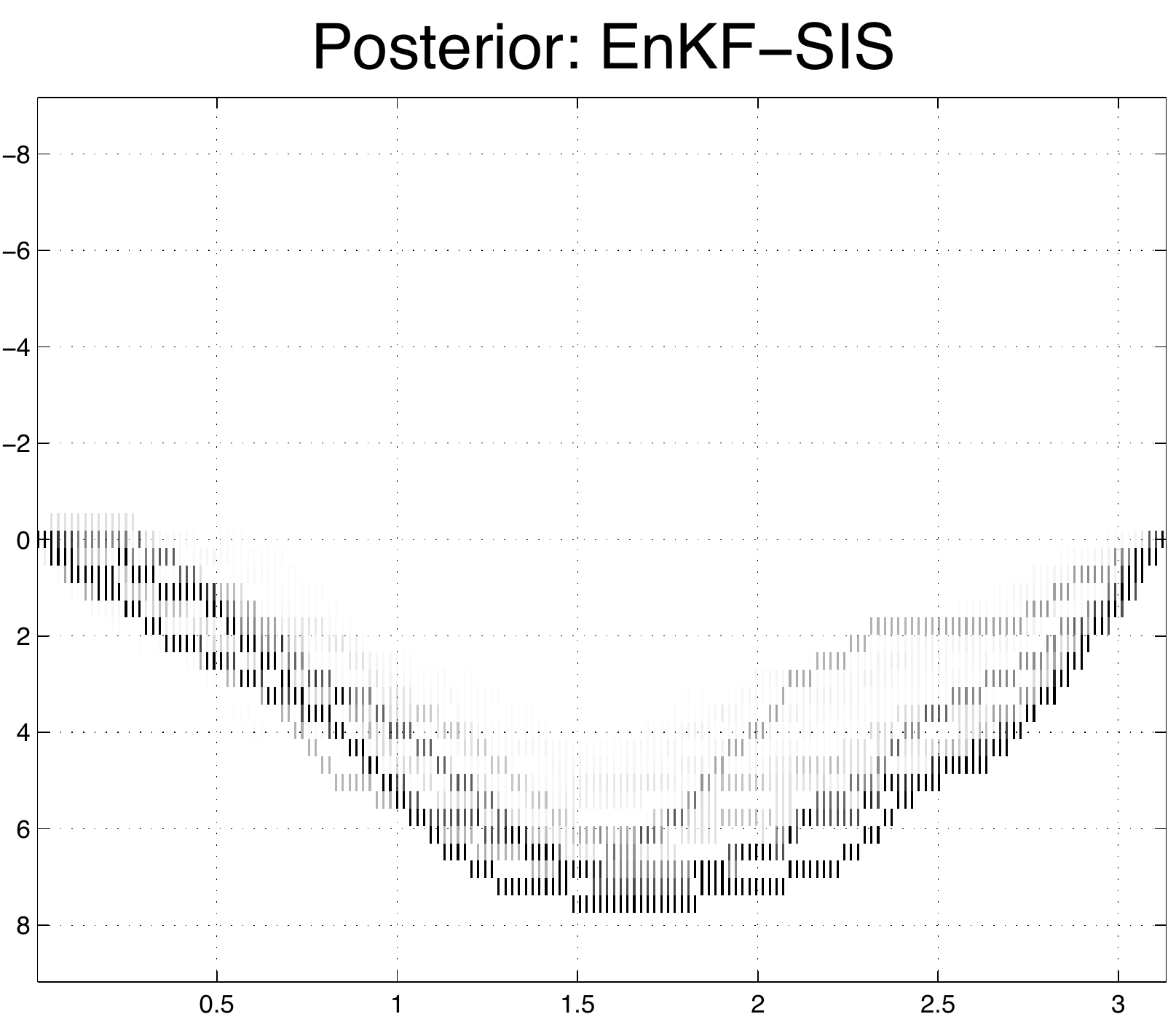}\\
{(c)} & {(d)}%
\end{tabular}
\end{center}
\caption{SIS cannot make a large update. The horizontal axis is the spatial
coordinate $x\in\lbrack0,\pi]$. The vertical axis is the value of function
$u$. The level of shading on each vertical line is the marginal density of $u$
at a fixed $x$, computed from a histogram with 50 bins. While EnKF and
EnKF-SIS create ensembles that are attracted to the data value $u(\pi/2)=7$,
SIS cannot reach so far because there are no such members in this relatively
small ensemble of size $N=100$. }%
\label{fig:sis-fail}%
\end{figure}

Finally, typical results for filtering in the space of functions on $[0,\pi]$
of the form%
\begin{equation}
u=\sum_{n=1}^{d}c_{n}\sin\left(  nx\right)  \label{eq:sin}%
\end{equation}
are in Figs.~\ref{fig:b-enkf} and \ref{fig:sis-fail}. The ensemble size was
$N=100$ and the dimension of the state space was $d=500$. The Fourier
coefficients were chosen $\lambda_{n}=n^{-3}$ to generate the initial ensemble
from (\ref{eq:random-smooth}), and $\kappa_{n}=n^{-2}$ for the norm in the
density estimation.

Fig.~\ref{fig:b-enkf} again shows that EnKF cannot handle bimodal
distribution. The prior was constructed by assimilating the data likelihood%
\[
p(d|u)=\left\{
\begin{array}
[c]{l}%
1/2\mbox{ if }u(\pi/4)\mbox{ and }\\
\quad\quad u\left(  3\pi/4\right)  \in\left(  -2,-1\right)  \cup(1,2)\\
0\mbox{ otherwise}
\end{array}
\right.
\]
into a large initial ensemble (size 50000) and resampling to the obtain the
forecast ensemble size $N=100$ with a non-Gaussian density. Then the data
likelihood $u\left(  \pi/2\right)  -0.1\sim N(0,1)$ was assimilated to obtain
the analysis ensemble.

Fig.~\ref{fig:sis-fail} shows a failure of SIS. The prior ensemble sampled
from Gaussian distribution with coefficients $\lambda_{n}=n^{-3}$ using
(\ref{eq:random-smooth}) and (\ref{eq:sin}), and the data likelihood was
$u\left(  \pi/2\right)  -7\sim N(0,1)$.

\section{Conclusion}

\label{sec:conclusion}

We have demonstrated the potential of a predictor-corrector filter to perform
a successful Bayesian update in the presence of non-Gaussian distributions,
large number of degrees of freedom, and large change of the state
distribution. Open questions include convergence of the filter in high
dimension when applied to multiple updates over time, mathematical convergence
proofs for the density estimation and for the Bayesian update, and performance
of the filters when applied to systems with a large number of different
physical variables and modes, as is common in atmospheric models.

\subsection*{Acknowledgements.}

This work was supported by the National Science Foundation under grants
CNS-0719641, ATM-0835579, and CNS-0821794.

%\bibliographystyle{splncs}
%\bibliography{../../bibliography/dddas-jm}

\end{document}